\documentclass[11pt,letterpaper]{article}

\usepackage{cvpr}
\usepackage{times}
\usepackage{epsfig}
\usepackage{graphicx}
\usepackage{amsmath}
\usepackage{amssymb}

\usepackage[pagebackref=true,breaklinks=true,letterpaper=true,colorlinks,bookmarks=false]{hyperref}

\cvprfinalcopy 

\ifcvprfinal\fi
\begin{document}

\title{An Analytical Approach for Insulin-like Growth Factor Receptor 1 and
Mammalian Target of Rapamycin Blockades in Ewing Sarcoma}

\author{Romeil Sandhu$^1$,  Salah-Eddine Lamhamedi-Cherradi$^2$,  Sarah Tannenbaum$^3$ \\ Joseph Ludwig$^2$, Allen Tannenbaum$^{3}$\\\\ \\
{\small [1] Department Biomedical Informatics, Stony Brook University}\\
{\small [2] Department of Sarcoma Medical Oncology, MD Anderson Cancer Center}\\
{\small [3] Division of Pediatric Hematology/Oncology/Stem Cell Transplantation, Columbia University Medical Center}\\
{\small [4] Departments Computer Science and Applied Mathematics \& Statistics, Stony Brook University}
}

\maketitle
\thispagestyle{empty}

\begin{abstract}
We present preliminary results that quantify network robustness and fragility of Ewing sarcoma (ES), a rare
pediatric bone cancer that often exhibits de novo or acquired drug resistance. By identifying novel proteins or
pathways susceptible to drug targeting, this formalized approach promises to improve preclinical drug development
and may lead to better treatment outcomes. Toward that end, our network modeling focused upon the
IGF-1R-PI3K-Akt-mTOR pathway, which is of proven importance in ES. The clinical response and proteomic networks
of drug-sensitive parental cell lines and their drug-resistant counterparts were assessed using two small
molecule inhibitors for IGF-1R (OSI-906 and NVP-ADW-742) and an mTOR inhibitor (mTORi), MK8669, such
that protein-to-protein expression networks could be generated for each group. For the first time, mathematical
modeling proves that drug resistant ES samples exhibit higher degrees of overall network robustness (e.g., the
ability of a system to withstand random perturbations to its network configuration) to that of their untreated or
short-term (72-hour) treated samples. This was done by leveraging previous work, which suggests that Ricci curvature,
a key geometric feature of a given network, is positively correlated to increased network robustness. More
importantly, given that Ricci curvature is a local property of the system, it is capable of resolving pathway fragility.
In this note, we offer some encouraging yet limited insights in terms of system-level robustness of ES and lay the
foundation for scope of future work in which a complete study will be conducted.
\end{abstract}

\section{Introduction}
\begin{figure}[!t]
\begin{center}
\includegraphics[height=7cm]{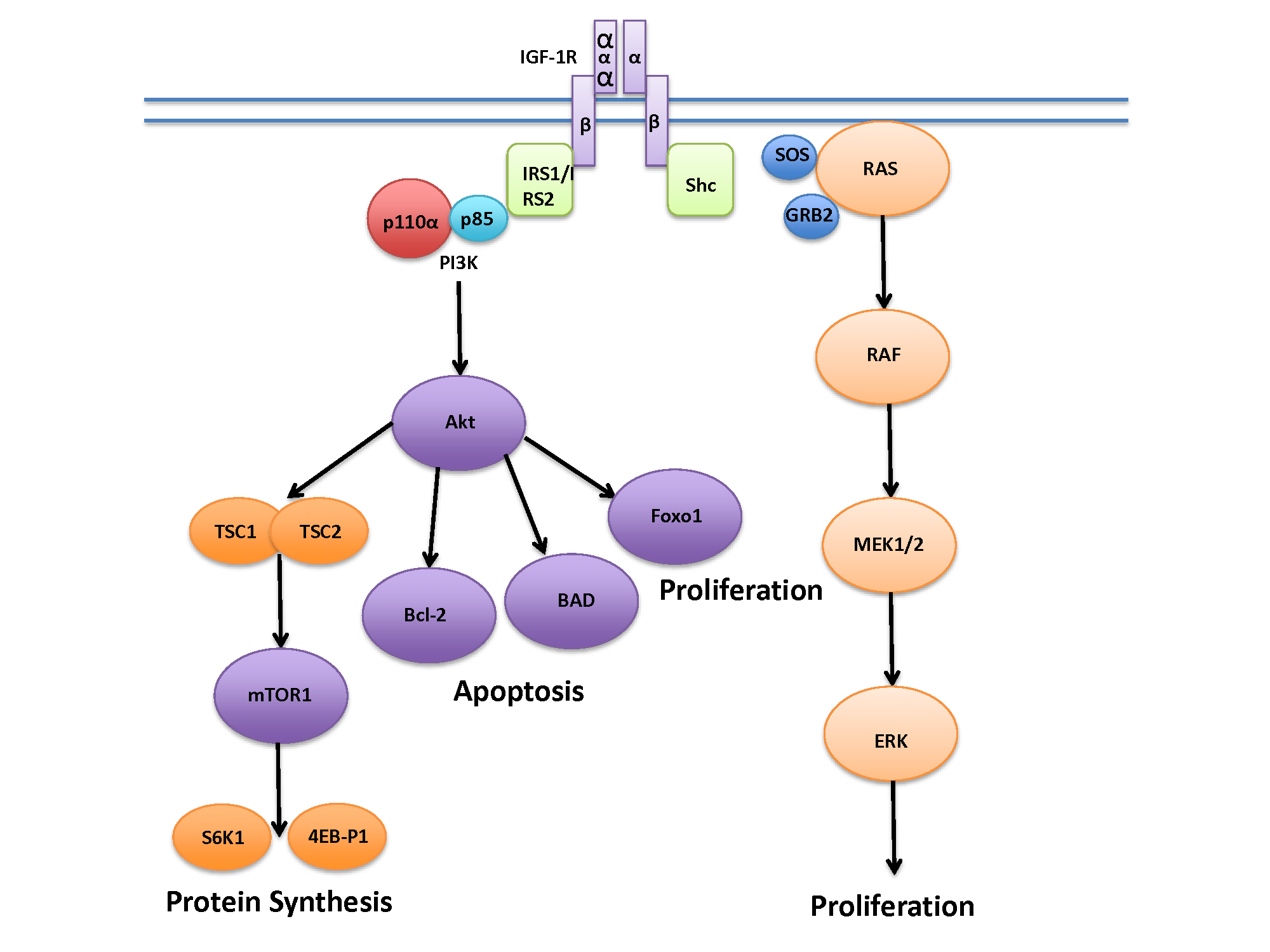}
\caption{A simple schema of the IGF-1R pathway.  IGF-1R is activated through binding of IGF-1 and IGF-2.  Autophosphorylation and crosslinking of IGF-1R leads to activation of IRS-1 and Shc and from there downstream activation of PI3K/Akt and Ras pathways.}
\label{fig:igfr1}
\end{center}
\end{figure}
Quantifying the fragility (and robustness) of biological pathways is a problem of paramount importance in guided targeted cancer therapy.  In this note, we restrict our attention to Ewing Sarcoma (ES) - which like many cancer types, often circumvents the effect of drug therapies due to alternative feedback signaling cascades.  In ES, compelling evidence has shown that the insulin-like growth factor 1 receptor (IGF-1R) is particularly important \cite {Kolb, Tolcher, Prieur, Mateo} as it is involved in the activation of multiple oncogenic pathways including the downstream activation of phosphoinositide-3-kinase/Akt/mammalian target of rapamycin (mTOR) \cite{Schwartz}.  To this end, studies have shown that targeting IGF-1R can inhibit growth in ES cells \emph{in vitro} and in xenograft models \cite{Scotland}; however, clinical success has been muted due to the the low response rate \cite{Kurzrock}.  As such, combinatorial inhibition of IGF-1R and mTOR has been shown to triple the response rate as compared to solely IGF-1R inhibition to limit inherent feedback \cite{Naing, Schwartz2, Atzori, Zhong}.  Although drug-resistant ES samples still arise, this noted synergistic effect with respect to duration of response lends support to notions of network ``fragility and ``robustness'' \cite{kitano}.  That is, on the one hand, the increase in response rate is correlated to an increase in fragility while, on the other hand, the formation of IGF-1R/mTOR drug resistance ES samples points to an increase in robustness (i.e., employing alternative feedback loops for continued metastasis).  This preliminary note lays the continued foundation to the following key question:
\begin{figure}[!t]
\begin{center}
\includegraphics[height=4cm]{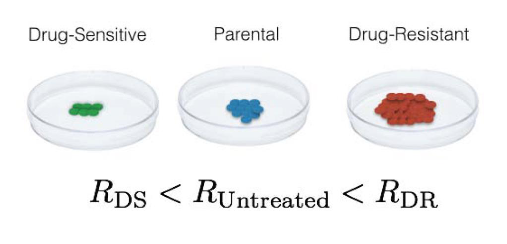}
\caption{Proposed Behavior of ES tumor.}
\label{fig:sweet_spot}
\end{center}
\end{figure}
\\
\\
\textbf{Central Question:}  \emph{Given the complexity and cost of selecting drug candidates, can we quantify (and therefore predict) pathway fragility in order to uncover a set of a set of n-tuple targets that can properly disrupt the alternative signaling cascades attributed to ES tumor growth?}
\\
\\
In addressing this critical question, a host of new ``-omics'' technologies has provided cancer biologists with
unparalleled access to expression data across the proteome or transcriptome that allows scientific exploration of
tumor phenotype at the network level, rather than a reductionist approach confined to specific protein-protein interactions. As such, one can consider cellular interactions apart of the interactome as a complex dynamical system represented by weighted graphs \cite{alon,barabasi,barabasi2,West}.  In doing so, we are able to compute certain geometric features in hopes of elucidating certain key attributes and stylized facts of the underlying system.  One such property is the geometric notion curvature \cite{DoCarmo}. We have previously shown that increases in Ricci curvature is positively correlated to increases in robustness, herein expressed as $\Delta Ric \times \Delta R\geq 0$ \cite{our_nature}. Roughly speaking, robustness relates to the rate at which a given dynamical system returns to its original (normal) state following a perturbation or external disturbance -  we refer the reader to \cite{kitano,Doyle, demo,dem} for a precise definition of robustness and several works \cite{LV, Romania,Mass,Zhou,Ollivier} regarding the details of curvature. Nevertheless, the key concept here is that on a given graph/interactome, we are able to employ Ricci curvature to express system-level and pathway fragility.  In the context of clinical success, it has been argued that feedback loops are essential to the function of ES with previous rationale for the success of joint IGF-1R/mTOR inhibition \cite{Schwartz}.  Stated in the context of graph theory, this feedback or the number of triangles in an interactome (redundant pathways) can be characterized by a lower bound of Ricci curvature \cite{our_nature, bauer, tetali}. We show some key pathways connected to IGF-R1 in Figure~\ref{fig:igfr1}.

We next note that we previously demonstrated that cancer networks exhibit a higher degree of global robustness (in terms of network curvature) in relation to their normal counterparts \cite{our_nature}.  We now focus our attention on comparing robustness for drug resistant, drug sensitive, and parental (untreated) samples specifically in Ewing Sarcoma (treated by OSI-906/NVP-ADW-742 and MK8669)  by constructing protein-to-protein correlation expression networks derived from a selected RPPA panel for each associative group.  We note that while topology may change under such longitudinal studies, for simplicity, we fix the topology of the networks using prior data on known physical interactions allowing only the weights to evolve. Then, by treating each network as a random walk, we attempt to exploit the underlying dynamics of specific protein-to-proteins interactions.  Nevertheless, we caution readers on the current preliminary results due to the limitation of sample size and scope of network pathways--\emph{ \textbf{this work is presented with the forethought that a larger exploratory analysis will be conducted to ensure clinical benefit}.}

\begin{figure*}[!t]
\begin{center}
\includegraphics[height=4cm]{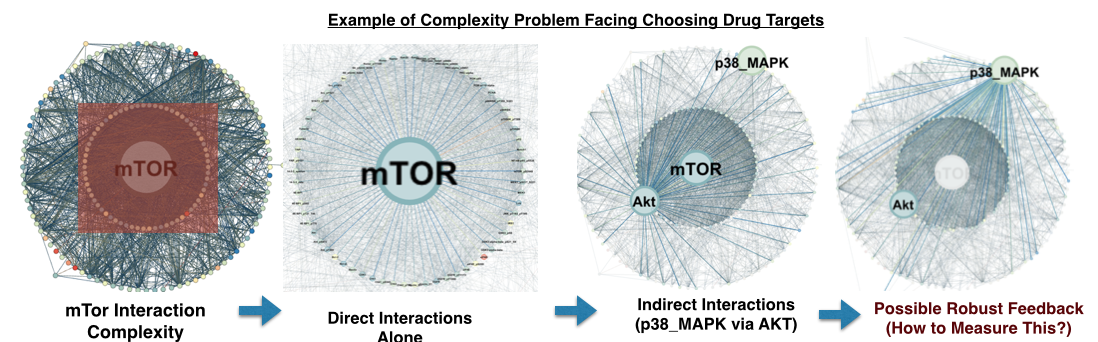}
\caption{An example of the complexity of choosing drug targets.  Only focusing on the mTOR signaling pathways, there exists a vast set direct interactions compounded further by the influence of indirect interaction (e.g., mTOR to p38\_MAPK via AKT).  The resulting structures builds various feedback (not seen in this figure) circumventing the effect of a particular therapy (e.g., series of interactions leading p38\_MAPK back to mTOR) .  The methodology proposed in this note attempts to provide a solution that can quantify pathway fragility and robustness of such indirect interactions.  In turn, we can then begin to understand mechanisms of resistance}
\label{fig:complexity_fig}
\end{center}
\end{figure*}
The remainder of the present note is outlined as follows: In the next section, we revisit a previously employed method for measuring network and pathway fragility via Ricci curvature, in the Ollivier sense \cite{Oll_markov,Ollivier}.  Section~\ref{sec:results} presents preliminary results demonstrating that resistant ES cell samples are more robust than their untreated counterparts with drug sensitive samples being the most fragile; this is seen in Figure \ref{fig:sweet_spot}.  We also elucidate possible proteins involved in pathways (e.g., MEK1, IRS1, mTOR, pS760K) that have been suggested as possible significance with regards to contribution to the robustness in ES.  We conclude with a discussion of future work.
\section{Measuring Fragility via Ricci Curvature}
\label{sec:fragility}
As mentioned in the previous section, we utilize the stylized fact that increases in Ricci curvature are positively correlated to increases in robustness, i.e., $\Delta Ric \times \Delta R\geq 0$.  Given that our work is confined to a discrete space of an interactome/graph in which the nodes on a graph represent a particular protein/gene and edges represent an ``interaction'', we employ a neat notion of a Ricci curvature \cite{Ollivier} inspired through coarse geometry.  In particular, if we let $(X,d)$ be a metric space equipped with a family of probability measures $\{\mu_x : x \in X\}$, we define the {\em Ollivier-Ricci curvature} $\kappa(x,y)$ along the geodesic connecting nodes $x$ and $y$ via
\begin{equation} \label{OR}
W_1(\mu_x,\mu_y) = (1-\kappa(x,y))d(x,y),
\end{equation}
where $W_1$ denotes the Earth Mover's Distance (Wasserstein 1-metric) \cite{Villani2,Villani3} and $d$ is the geodesic distance on the graph.
For the case of weighted graphs, we set
\begin{eqnarray} d_x &=& \sum_y w_{xy}\\
\mu_x(y)&:=& \frac{w_{xy}}{d_x} ,\end{eqnarray}
where $d_x$ is the sum taken over all neighbors of node $x$ and where $w_{xy}$ denotes the weight of an edge connecting node $x$ and node $y$ ($w_{xy}=0$ if there is no direct edge connecting node x to node y). The measure $\mu_x$ may be regarded as the distribution of a one-step random walk starting from $x$, with the weight $w_{xy}$ quantifying the strength of interaction between nodal components or the diffusivity across the corresponding link (edge).  Computationally, the above term $W_1$ may be computed as a linear program \cite{Rubner} allowing for several numerical advantages \cite{our_nature}. From this, we can then measure the fragility of a given (not necessarily adjacent) pathway through decreases/increases in the term $\kappa(x,y)$ with the hopes that such measures will be able to guide drug targeting therapies through such biological complexity.  An example of this complexity can be seen in Figure \ref{fig:complexity_fig}.
\section{Results}
\label{sec:results}
In this section, we present details on the data experimentally obtained as well the resulting computation for network curvature/robustness.
 \begin{table*}[ht]
     \begin{center}
     \begin{tabular}{  c }
\includegraphics[height=2.7cm]{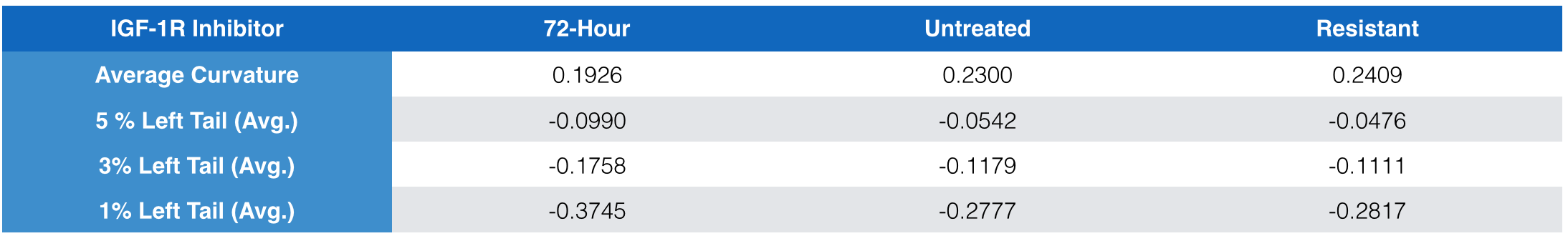}

      \end{tabular}
      \caption{Global network results for IGF-1R treatment show that, on average, robustness as measured by Ricci curvature is higher in drug resistant samples with drug sensitive samples being the most fragile. The analysis presented in this table shows both average Ricci curvature, minimum curvature, and average value of the left tails (at specific lengths) in order to characterize the ``shifts'' seen within each network.}
      \label{table:IGF_global}
      \end{center}
      \end{table*}
\subsection{Data and Network Construction}
The preliminary data analyzed was obtained from MD Anderson investigating mechanisms of resistance in TC32 and TC71 ES cell lines focusing on IGF-1R and/or mTOR inhibition; see \cite{salah_paper} for complete experimental details.  In short, two independent methods were used to generate drug resistance in ES cell lines grown in monolayer culture. In the first method for analyzing IGF-1R inhibition, parental cell lines were exposed to steadily increasing drug concentrations of OSI- 906 or NVP-ADW-742 for 7 months; herein referred to as \textbf{drug resistant}. In the second method, ES cells were exposed to exceedingly high concentrations for 72 hours before viable clones were expanded; herein referred to as \textbf{drug sensitive}.  These two groups were combined along a set of parental samples; herein referred to as \textbf{untreated / 72-Hour}.  A separate, but similar experiment was repeated with MK8669 to generate a complementary data set targeting the effects of mTOR inhibition. All together, there consists a set of 12 drug resistant samples, 4 drug sensitive samples, and 8 parental (untreated) samples for the OSI-906/NVP-ADW-742 (IGF-1R) treatment while 8 drug resistant samples, 2 drug sensitive samples, and 3 parental samples for the MK8669 (mTOR) treatment were obtained.  Using the protein expression data obtained from the RPPA panel (consisting of a small set of 112 and 165 proteins for IGF-1R and mTOR experiments, respectively), we were able to generate a protein-to-protein network through known physical interactions and with weights that were computed from the correlation across such samples within each group.  We should note that although the OSI-906/NVP-ADW-742 target IGF-1R inhibition, the data examined does not include IGF-1R expression and lacks several proteins (due to experimental reasons conducted outside the scope of the present work).  However, the data analyzed for MK8669 treatment did include mTOR and several significant interactions/proteins in this pathway. These data limitations will be a subject of future work.  Nevertheless, six ES networks characterizing the drug resistant, drug sensitive, and untreated cases for IGF-1R and separately, mTOR inhibition, were constructed and were sufficient for macroscopic analysis.
\subsection{Global Network and Pathway Fragility}
Following our previous work \cite{our_nature}, we conducted a global analysis with respect to network curvature on all six networks to highlight the inherent robustness in ES resistant cell lines.  In particular, Table \ref{table:IGF_global} and Table \ref{table:mtor_global} present the average Ricci curvature in each of the three groups for samples treated with NVP-ADW-742/OSI-906 and MK8669 as well as average value of the left tails (at given lengths of 5\%, 3\% and 1\%).  As such, one can see that the difference between drug resistant and untreated (and/or drug sensitive) samples is positive signifying robustness.  Moreover, the drug sensitive samples exhibit fragility when compared to the untreated and resistant cases, providing further evidence to not only the robust nature of metastasis but also the effect of drug treatments.  In addition to the average statistic, one can consider the left tail of the distribution as the lower bound of Ollivier-Ricci curvature as opposed to simply taking minimum value which maybe sensitive to topological errors.   The motivation for analyzing the lower bound / left tails of the distribution is significant as it is intimately connected to a notion of entropy which also serves as a proxy for robustness \cite{our_nature, LV}.  Again, we see that increases in the average value of tails is in line with the above analysis.

 \begin{table*}[ht]
     \begin{center}
     \begin{tabular}{  c }
\includegraphics[height=2.7cm]{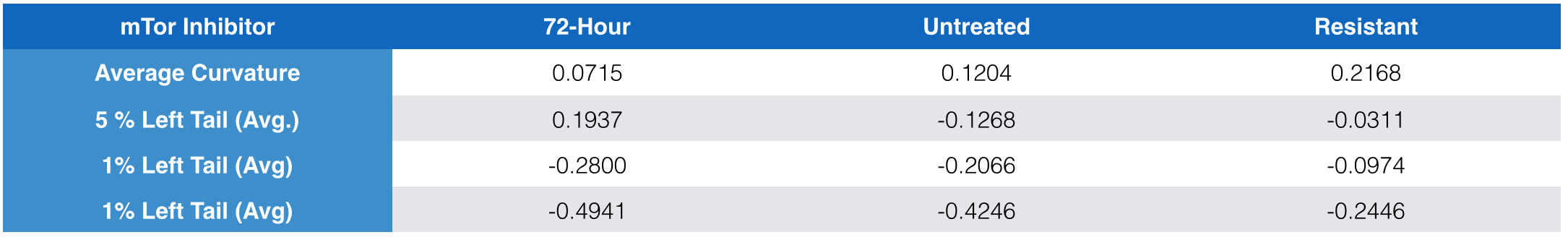}

      \end{tabular}
      \caption{Global network results for mTOR inhibition via MK8669 also shows that, on average, robustness as measured by Ricci curvature is higher in drug resistant samples with drug sensitive samples being the most fragile. The analysis presented in this table shows both average Ricci curvature, minimum curvature, and average value of the left tails (at specific lengths) in order to characterize the ``shifts'' seen within each network.}
      \label{table:mtor_global}
      \end{center}
      \end{table*}
While global results provide a quantitative indicator of network robustness/fragility,  the clinical benefit of the above method is that the methodology measures fragility in pathways (that need not be necessarily directly connected) for particular targets.  We can analyze particular targets in varying manners, one of which is through scalar curvature \cite{our_nature}.  Specifically, we define scalar curvature in this text as
\begin{equation}
S(x) = \sum_y \kappa(x,y)
\end{equation}
which is simply the sum all Ricci curvature over direct and indirect pathways.  The decision to employ this definition as opposed to simply summing over direct interactions is due to the fact that we would like to measure activity of a particular protein with respect to all other proteins in the context of robustness/fragility.   Thus, unlike differential expression, scalar curvature defined in this manner measures \textbf{\emph{hidden effects}} due to the underlying network. To this end, Table \ref{table:IGF_scalar} and Table \ref{table:mTor_Scalar} present several different proteins and their scalar curvature under drug resistant, drug sensitive, and untreated cases.
 \begin{table}[ht]
     \begin{center}
     \begin{tabular}{  c }
\includegraphics[height=3.5cm]{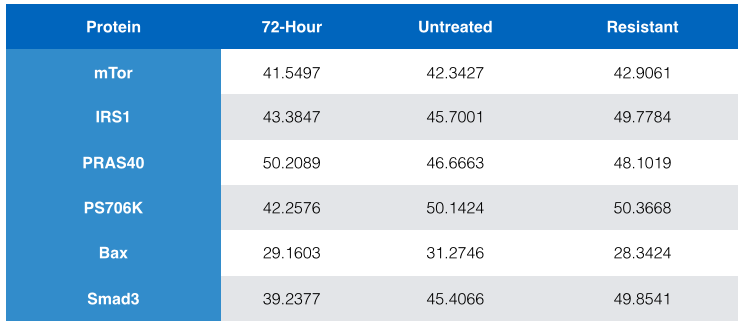}

      \end{tabular}
      \caption{Scalar curvature results for several possible targets in Ewing Sarcoma when samples were treated with OSI- 906 or NVP-ADW-742, which are known to inhibit IGF-1R/IR-$\alpha$ and IGF-1R.}
      \label{table:IGF_scalar}
      \end{center}
      \end{table}

\begin{figure*}[!t]
\begin{center}
\includegraphics[height=4cm]{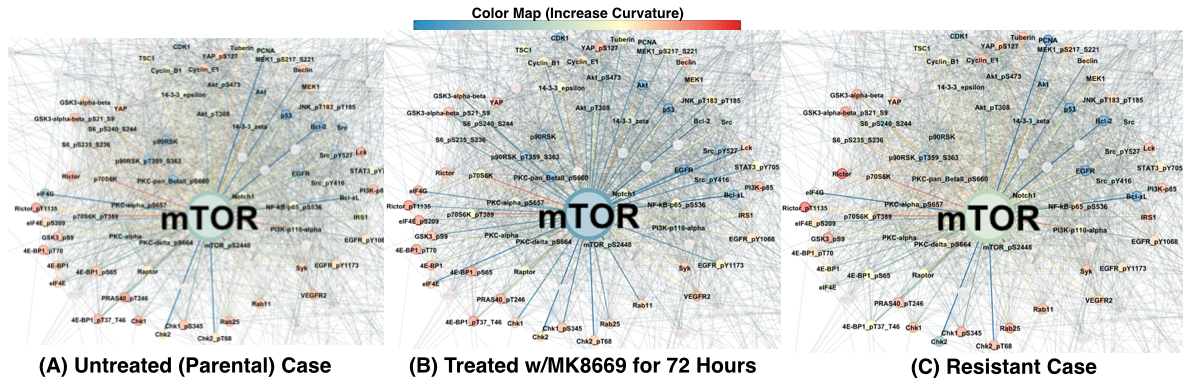}
\caption{This illustrates the increase in fragility of mTOR and the surrounding ``neighborhood'' as a result of the treatment of MK8669 for 72-hours.  Unfortunately, resistance becomes to build over time muted the the therapeutic effect of solely inhibiting mTor.}
\label{fig:mtor_treated}
\end{center}
\end{figure*}

Interestingly, we observe in Table \ref{table:IGF_scalar} those direct/indirect pathways involved with particular proteins, i.e., IRS1, that are in the ``neighborhood of'' of IGF-1R and IR-$\alpha$ elicit an increase to their robustness from drug sensitive samples to drug resistant samples. On the other hand, mTOR seems, for the most part, unaffected by OSI-906/NVP-ADW-742 treatment as compared to the relative change in Table \ref{table:mTor_Scalar}.  As seen in Figure \ref{fig:mtor_treated}  This direct quantitative effect from the treatment of MK8669 with regards to mTOR fragility is what one would expect when targeting a particular protein.

Moreover, several other insights are noted.  In particular, treatment with MK8669 responds to an increase activation of IRS1 as opposed to OSI-906/NVP-ADW-742.  Another protein, PRAS40, apart of the mTOR complex I (mTORC1), elicited an increase in fragility when treated with MK8669.  Biologically, this falls in line with previous understanding of mTORC1 sensitivity to such targeting.  The opposite effects were registered in IGF-1R samples providing further (yet very limited) evidence that mTOR serves as a powerful feedback structure when solely inhibiting IGF-1R \cite{Schwartz}.  In short, while such results are very preliminary, much of the biological understanding seen in previous literature aligned with our analysis on curvature on this small set of samples.

\section{Future Work Summary}
\label{sec:future_work}
In the above analysis, there exist several limitations that will be addressed.  First and foremost, the sample size and RPPA panel of proteins are quite small and for any clinical benefit to arise, we must extend the above analysis.  Of course, given the high costs associated with such targeted experiments, we are still able to uncover encouraging results that will help lay the foundation for a quantitative framework.  Additionally, given a complete set of pathways (RPPA panel), the next step will be to address specific pathways in the context of robustness and fragility.  Nevertheless, this note has provided additional evidence that drug resistant samples are more robust than untreated/drug sensitive samples.

In future work, we would also like to consider the application of curvature to a key issue in epigenetics. More precisely, the classic metaphor of Conrad Waddington \cite{Waddington1957} is based on a representation of epigenetic cellular differentiation as an ``epigenetic landscape,'' in which one imagines a downhill progression from an undifferentiated stem or progenitor cell state to a physiologically mature one. Thus the image is one of cells rolling down a ``potential energy hill'' into ``valleys'' that correspond to the final determination of the cellular states. Relative to this landscape, dedifferentiation refers to the process in which cells travel back up their differentiation path, through the epigenetic landscape, to become more immature and finally convert into the pluripotent state. This is really the concept underpinning ``cellular reprogramming'' in the cancer literature; see \cite{Zhang2013} for an interesting discussion about the reprogramming of sarcoma cells as well as \cite{Furusawa2012} for a dynamical systems perspective of the Waddington model.

In order to flush out the metaphor, one needs precise notions of the potential energy driving this process of cellular differentiation. In some very nice work \cite{Banerji2013}, the authors propose the use of entropy as the measure. As the authors note, recent work has demonstrated that local signaling entropy may serve as a novel indicator of drug sensitivity, while at the same time, may operate as a proxy for the height or elevation in Waddington's differentiation landscape. Furthermore, it has been argued that feedback loops are essential to the function of biological mechanisms and systems that arise from deliberate Darwinian-like principles \cite{Doyle, kitano}. Along these lines, Ollivier-Ricci curvature in addition to measuring robustness, can be regarded as a new feedback measure, i.e., the number of triangles in a network (redundant pathways) can be characterized by a lower bound of Ricci curvature [Bauer2011]. It is positively correlated to entropy and thus may play the role of epigenetic ``potential energy.'' It also has some computational advantages, is less sensitive to noise in the data than entropy, and gives both local and global information \cite{our_nature}. This key interplay between feedback, robustness, entropy, and network geometry as represented by curvature lies at the foundational heart of this part of our future research. The concept of feedback control has certainly been considered in various papers in the cancer literature. For example, very relevant to the work described in the present proposal is the research on the role of EWS-FLI1 in the inhibition of cytoskeletal autoregulatory feedback  \cite{Katschnig2017}.

Along these lines, understanding the dynamics of epigenetic regulation and gene program control are essential to unraveling the underlying mechanisms of epigenetic deregulation for solid tumors as described in \cite{Lawlor2015}. The geometric network approach using ES as a model cancer and EWS-FLI1 as a model oncogene/protein can of course be very relevant in such an endeavor.

\begin{table}[ht]
     \begin{center}
     \begin{tabular}{  c }
\includegraphics[height=3.5cm]{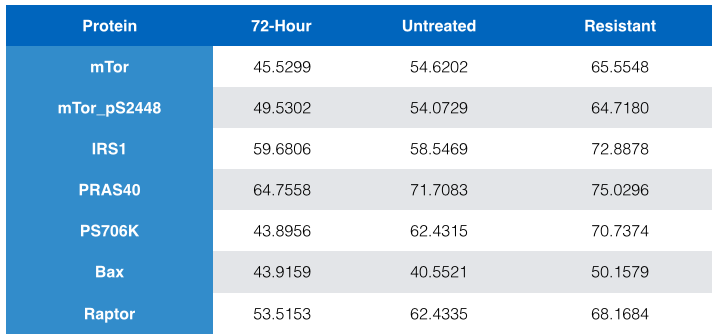}

      \end{tabular}
      \caption{Scalar curvature results for several possible targets in Ewing Sarcoma when samples were treated with MK8669, which is known to inhibit \emph{mTOR}.}
      \label{table:mTor_Scalar}
      \end{center}
      \end{table}

\section*{Acknowledgements}
This study was supported by AFOSR grants (FA9550-17-1-0435, FA9550-20-1-0029), NIH grant (R01-AG048769), and a grant from Breast Cancer Research Foundation (grant BCRF-17-193).

\end{document}